\documentclass[draftclsnofoot, onecolumn, comsoc, 12pt]{IEEEtran} 

\usepackage{graphicx}
\usepackage[noadjust]{cite}
\usepackage{mcite}
\usepackage{amsfonts,helvet}
\usepackage{fancyhdr}
\usepackage{threeparttable}
\usepackage{epstopdf}
\usepackage{amsthm}
\usepackage{amsmath}
\usepackage{siunitx}
\usepackage{amssymb}
\usepackage{dsfont}
\usepackage{color}
\usepackage{algorithm}
\usepackage{algpseudocode}
\usepackage{algcompatible}
\usepackage{enumerate}
\usepackage{gensymb}
\usepackage{cancel}
\usepackage{eucal}

\usepackage{boldline}
\usepackage{footnote}
\usepackage[hang,flushmargin]{footmisc}
\usepackage{caption}
\usepackage{subcaption}

\makeatletter
\newcount\SOUL@minus
\makeatother

\begin{document}

\title{Coordinated Rate-Splitting Multiple Access\\ for Integrated Satellite-Terrestrial Networks \\with  Super-Common Message }

%\author{Juhwan Lee,~\IEEEmembership{Student Member,~IEEE},  Jungwoo Lee,~\IEEEmembership{Senior Member,~IEEE},\\ Wonjae Shin,~\IEEEmembership{Senior Member,~IEEE}, and Bruno Clerckx,~\IEEEmembership{Fellow,~IEEE} 
\author{Juhwan Lee,  Jungwoo Lee, Longfei Yin, Wonjae Shin, and Bruno Clerckx
\thanks{Juhwan Lee and Jungwoo Lee are with the Department of Electrical and Computer Engineering, Seoul National University, Seoul 08826, South Korea (e-mail: {\texttt{\{sgsyk649, junglee\}@snu.ac.kr}}). %\texttt{jhlee@cml.snu.ac.kr} Gwanak-gu, Seoul 08826,
Longfei Yin is with the Department of Electrical and Electronic Engineering, Imperial College London, London SW7 2AZ, U.K. (e-mail: \texttt{longfei.yin17@imperial.ac.uk}). 
Wonjae Shin is with the Department of Communications Engineering, Korea University, Seoul 02841, South Korea (e-mail: \texttt{wjshin@korea.ac.kr}). %Suwon 16499
Bruno Clerckx is with the Department of Electrical and Electronic Engineering, Imperial College London, London SW7 2AZ, U.K. and with Silicon Austria Labs (SAL), Graz A-8010, Austria (e-mail: \texttt{b.clerckx@imperial.ac.uk%; bruno.clerckx@silicon-austria.com
}) (\textit{Corresponding author: Wonjae Shin}). 
%This work was supported in part by the National Research Foundation of Korea (NRF, 2021R1A2C2014504, 2021R1A4A1030775, 2022R1A2C4002065), Institute of Information \& Communications Technology Planning \& Evaluation (IITP, 2021-0-00106, 2021-0-01059, 2021-0-00260, 2022-0-00704), Samsung Electronics Co., Ltd (Contract ID: MEM210728\_0001), INMAC, and BK21 FOUR program.
} 
% the Communications and Signal Processing Group / London SW7 2AZ
%\thanks{Manuscript received XXX, XX, 2022; revised XXX, XX, 2022.}

}
%\markboth{IEEE Transactions on Vehicular Technology,~Vol.~XX, No.~XX, XXX~2023}
{}
%{Shell \MakeLowercase{\textit{et al.}}: Bare Demo of IEEEtran.cls for Journals}

\maketitle

\begin{abstract}
Rate-splitting multiple access (RSMA) is an emerging multiple access technique for multi-antenna networks that splits messages into common and private parts for flexible interference mitigation.
Motivated by its robustness and scalability, it is promising to employ RSMA in integrated satellite-terrestrial networks (ISTN), where a satellite serves satellite users (SUs) broadly with a multibeam multicast transmission while terrestrial base station (BS) serves cellular users (CUs) with a unicast transmission, operating in the same frequency band.
To avoid the data exchange between satellite/cellular networks via backhaul, we assume a coordinated ISTN relying  on imperfect channel state information.
%with no tight synchronization at the data level across the satellite and BSs. 
%
We put forth a  coordinated RSMA  framework tailored to the coordinated ISTN by applying {\it inter-network} rate-splitting (RS) with a super-common message on top of {\it intra-network} RS with common/private messages.
With the unified RS design for inter- and intra-networks, we jointly optimize the precoding and power allocation of the private/common/super-common messages to achieve max-min fairness among all SUs and CUs through successive convex approximation.
By doing so, the power of the super-common message can be adjusted according to interference levels of the satellite towards CUs, thereby potentially mitigating inter-network interference. 
%by which performance gains from successive interference cancellation (SIC) are attainable
%With the joint rate-splitting in inter/intra-network, the power is able to be allocated to split messages adaptively on the corresponding link levels.
Simulation results demonstrate the superiority and robustness of our approach to cope with various interference and propagation conditions.
\end{abstract}
\vspace{10mm}


\IEEEpeerreviewmaketitle

\section{Introduction}

With advances in satellite technology and the growth of data demand, integrated satellite-terrestrial networks (ISTN) have attracted considerable attention due to their potential to provide reliable and ubiquitous coverage over large geographic areas. Satellites can expand broadband coverage into rural or far remote areas where terrestrial networks have limited coverage, thereby complementing terrestrial networks [\ref{ref:Access_Survey}].
In the ISTN, a multibeam satellite widely serves satellite users (SUs) with a multicast transmission, while a terrestrial base station (BS) serves cellular users (CUs) with a unicast transmission. 
Across the multiple spot beams and cells, higher throughput can be achieved through full frequency reuse [\ref{ref:Magazine_Signal Processing for HTS}].
%Through the multibeam multicast transmission, higher throughput can be achieved via full frequency reuse across multiple spot beams [\ref{ref:Magazine_Signal Processing for HTS}]. 
With the aggressive frequency reuse, however, severe interference can be induced in and between satellite networks and cellular networks. Moreover, it is difficult to acquire accurate channel feedback at the gateway (GW) in a coherence time, particularly due to the fast-moving of satellite and high end-to-end propagation delay, following channel state information (CSI) uncertainty.
% sub-network 정의가 없으니 헷갈림, 
%Moreover, it has high end-to-end propagation delay and it is hard to track the channel variation due to the rapid motion of satellite, thus it is difficult to get accurate channel feedback at GW in a coherence time, following channel state information (CSI) uncertainty.

Rate-splitting multiple access (RSMA) has recently emerged as a promising transmission scheme to manage interference [\ref{ref:2019 Letter_Bruno_RSMA}], [\ref{ref:2018 EURASIP_Bruno_RSMA}] and to improve robustness to imperfect CSI at the transmitter (CSIT) in multi-antenna multi-user networks [\ref{ref:2016 TCOM_Bruno_RSMA imperfect CSIT}]-[\ref{ref:2016 TSP_Bruno_RSMA imperfect CSIT}]. 
%RSMA 설명에 SDMA, NOMA도 추가하면서 그 뒤에 flexibility 언급하기
For the RSMA transmission, at the transmitter, the message for each user is split into common and private parts. 
All common parts are combined and encoded into a common stream, while private parts are individually encoded into private streams. The users decode and remove the common stream through successive interference cancellation (SIC).
After the SIC, each user decodes the intended private stream while treating the other users' signals as noise. 
%Throughout this process, RSMA partially decodes interference and partially treats interference as noise. 
As such, RSMA can generalize space-division multiple access (SDMA) where interference is fully treated as noise, and non-orthogonal multiple access (NOMA) where interference is fully decoded.
Based on this flexibility,  RSMA has been utilized in inter-beam interference managements for multibeam satellite communications, resulting in  more robust to CSIT inaccuracies \mbox{[\ref{ref:TCOM_RSMA for Multibeam Satellite}]-[\ref{ref:2023 Magazine_RSMA for 6G}]}.

The superiority of RSMA for managing interference in ISTN has been investigated in [\ref{ref:Arxiv_Bruno_RSMAinISTN}] by proposing a joint beamforming design to achieve the max-min fairness (MMF).
%MMF 가 중요한 metric인 이유 in ISTN (coverage expansion, sum-rate는 특정 user)
The MMF rate is an important metric in the ISTN in terms of ubiquitous coverage, while sum-rate maximization can be achieved by allocating all the power to only a few users with the best channel conditions.
%at the expense of users with poor channel condition.
{In cooperative ISTN, the transmit data is fully shared across the satellite and BS at the GW; however, this can incur significant signaling overhead. 
On the other hand, in coordinated ISTN, the satellite and BS do not share their data, thereby not requiring tight synchronization at the data level across the satellite and BS.} 
In RSMA-based coordinated ISTN  \mbox{[\ref{ref:Arxiv_Bruno_RSMAinISTN}]}, inter-beam interference and intra-cell interference are partially decoded; 
however, it only considers the rate-splitting (RS) in each intra-network within the satellite network and cellular network separately, while inter-network interference is fully treated as noise. 
%the framework proposed in \mbox{[\ref{ref:Arxiv_Bruno_RSMAinISTN}]}
{Besides, it only deals with a geostationary (GEO) satellite scenario with an altitude of 36,000 km above the surface of the Earth, which suffers from a low level of inter-network interference; however, it inherently has the disadvantage of high launch costs, poor link budget, and long propagation delay.} 
Along with \mbox{[\ref{ref:Arxiv_Bruno_RSMAinISTN}]}, the existing studies on multi-cell RSMA transmission mainly resort to i) treating inter-cell interference as noise \mbox{[\ref{ref:Arxiv_Bruno_RSMAinISTN}]-[\ref{ref:Multicell-RSMA_2}]}, or ii) inter-cell RS with data sharing between cells \mbox{[\ref{ref:Multicell-RSMA_3}]-[\ref{ref:Multicell-RSMA_5}]}. 
Different from the previous studies, to manage both intra- and inter-network interference without data sharing, our framework is motivated by the Han-Kobayashi scheme \mbox{[\ref{ref:Han Kobayashi}]}, which is the quasi-optimal transmission strategy for two-user Gaussian interference channel. 
In the Han-Kobayashi scheme, power allocation to common/private messages is flexibly determined according to relative channel gains between the desired and interference links. 
%a different power is allocated to common/private messages according to relative channel gains between the desired and interference links. 
Inspired by this, we propose a coordinated RSMA framework by applying inter-network RS with a super-common message on top of intra-network RS with common/private messages. 
With the unified RS design for inter- and intra-networks, we jointly optimize the precoding and power allocation of the private/common/super-common messages to achieve the MMF among all users.
%that enables to partially decode the residual interference rather than fully treating it as noise to improve the MMF rate.
% which can partially decode the residual interference rather than fully treating as noise.
% Interference를 fully noise 취급하지 않고 partially decode 할 수 있는 RSMA scheme propose
% Han-Kobayashi에서 hint얻었다. (refer추가)
%In particular, we investigate a coordinated RSMA transmission framework for the coordinated ISTN, which jointly performs rate-splitting in inter-network with super-common message as well as in intra-network.
By doing so, the transmit power of the super-common message can be adjusted according to the interference levels of the satellite towards CUs, thereby  mitigating inter-network interference. 
{Thus, it can also be effectively applied in emerging ISTN scenarios using low Earth orbit (LEO) satellites with an altitude of less than 2,000 km, since its higher level of inter-network interference can be subtracted thanks to the use of super-common message.}
%In addition, from the GEO satellite scenario, we further consider up to a low Earth orbit (LEO) satellite scenario with an altitude of less than 2000 km, which has a high level of inter-network interference as well as interference in the satellite network. 
To the best of our knowledge, there are no existing studies that attempt to manage both inter- and intra-network interference for coordinated ISTN without data sharing among networks in the context of RSMA. 
%We utilize idea of the Han-Kobayashi scheme, where the power is allocated to the split messages according to the relative channel gains between the desired link and interference [\ref{ref:Han Kobayashi}].
%which additionally employs a super-common message at satellite, which can partially decode the interference between satellite and cellular networks, improving the MMF rate.
%enabling to partially decode interference between the satellite and cellular networks.
%as well as satellite-common message at satellite and cellular-common message at BS.
% adopt 단어 수정 (새로운 거를 함으로써 어떻게 되는지, 성능을 좋게 한다는 이야기)
The contributions of this paper are summarized as follows:
\begin{itemize}
%\vspace{-0.5mm}
\item We employ a super-common message that can be initially decoded and deducted from all users through SIC.
Through this novel RSMA framework, the interference between the satellite network and cellular network can be flexibly decoded and  treated as noise.% according to interference conditions.
%Through flexible power splitting between split messages adaptively at the level of interference in and between satellite-network and cellular-network, the interference can be partially decoded and partially treated as noise. 
%\item \hl{With limited knowledge of satellite channels, we formulate an optimization framework to achieve the MMF among all users. We approximate the problem into a more tractable one via successive convex approximation (SCA).}
\item {With limited knowledge of satellite channels, we formulate an optimization framework to achieve the MMF among all users, using a concept of generalized mutual information (GMI). 
We approximate it into a more tractable one via successive convex approximation (SCA).}
\item We numerically demonstrate the superiority and CSI-error robustness of our proposed framework, compared to conventional  RSMA, SDMA, and OMA counterparts. 
{Through this, the proposed scheme is shown to have the potential to provide substantial gain, especially when a satellite is operated in LEO and the power budget of a terrestrial BS is high enough.}
\end{itemize}

%The rest of this paper is organized as follows. The system model is described in Section \uppercase\expandafter{\romannumeral2}. \uppercase\expandafter{\romannumeral3} shows the MMF problem formulation and the design of a CSI-robust joint beamforming scheme. Numerical results are presented in \uppercase\expandafter{\romannumeral4}. Finally, \uppercase\expandafter{\romannumeral5} concludes this paper.

%Throughout this paper, the notation of $\mathbb{C}$, $\mathbb{E}\{ \cdot \}$, $\lVert \cdot \rVert$, $ \left\vert \cdot \right\vert$, $(\cdot)^{T}$, $(\cdot)^{H}$, $\circ$, $J_{k}$ denote complex space, expectation, absolute value, Euclidean norm, transpose, Hermitian transpose, Hadamard product and first-kind Bessel function with order $k$, respectively.
%\vspace{-4.5mm}
\section{System Model}
%\vspace{-1mm}
We consider a coordinated ISTN system with a single GW, satellite, terrestrial BS, and single-antenna equipped multiple SUs and CUs. %as described in Fig.~\ref{fig:System Architecture}. %where the satellite and terrestrial BS do not share the transmitted data information at the GW. 
In the coordinated ISTN, the CSI of all links is shared between the satellite and BS at the GW, while the transmit data is not shared. 
To enhance spectral efficiency, the frequency band is fully reused among all users. 
%To enhance the spectral efficiency, full frequency reuse is considered in which all SUs and CUs operate in the same frequency band. 
A single feed per beam (SFPB) architecture with array fed reflector antenna is assumed, generating $N_{s}$ adjacent beams with $N_{s}$ antenna feeds. 
%(i.e. one feed per one beam). 
The beams are indexed by $\mathcal{N}_{s} \triangleq \{1,...,N_{s}\}$. 
There are $\rho$ users within each beam, and totally $K_{s}=\rho N_{s}$ SUs. 
For the satellite network,  multibeam multicast transmission is implemented based on DVB-S2X technology \mbox{[\ref{ref:DVB-S2X}]}, in which multiple SUs of each beam are served simultaneously with a single coded frame.
%\hl{For the satellite network, multibeam multicast transmission is implemented, in which multiple SUs of each beam are served simultaneously with a single coded frame based on DVB-S2X technology }[\ref{ref:DVB-S2X}]\hl{.
%
%, in which a single coded frame based same information is simultaneously decoded by the SUs in each beam.
%Following the DVB-S2X technology, the satellite employs multibeam multicast transmission in which a same information is simultaneously decoded by all SUs in each beam.
The feeder link between the GW and the satellite is assumed to be noiseless, and the link between the GW and the terrestrial BS is assumed to be an ideal optical link. 
Moreover, we assume that the BS equipped with $N_{t}$ antennas serves $K_{t}$ unicast CUs which are densely populated with  $K_{t} \leq N_{t}$. 
The CUs are indexed by $\mathcal{K}_{t} \triangleq \{1,...,K_{t}\}$. Assuming that the SUs are located out of the BS's service area, the channel between the BS and all SUs is negligible. 
The power budget at the satellite and the BS are respectively $P_{s}$ and $P_{t}$.
%As shown in Fig.~\ref{fig:System Architecture}, 
The GW is a central controller for resource allocation, interference management, collecting data information, and feeding both satellite and BS.
%\vspace{-10mm}
%\vspace{-5mm}
\subsection{Channel Model}
%\vspace{-1mm}
%\vspace{-0.5mm}
The channel model in the ISTN system is composed of the following satellite and terrestrial channel models.

\subsubsection{Satellite Channel Model}
The satellite channel matrix between the satellite and all SUs is characterized as $\mathbf{F}=[\mathbf{f}_{1},...,\mathbf{f}_{K_{s}}]\in \mathbb{C}^{N_{s}\times K_{s}}$, where $\mathbf{f}_{k_{s}}$ denotes the channel between the satellite and SU-$k_{s}$. 
We assume that GW cannot accurately estimate the satellite CSI. 
%Given the assumption of imperfect CSI at GW, 
Thus, the channel $\mathbf{f}_{k_{s}}$ is formulated as $\mathbf{f}_{k_{s}} = \hat{\mathbf{f}}_{k_{s}} + \mathbf{e}_{k_{s}}$, where the channel estimate $\hat{\mathbf{f}}_{k_{s}}$ is expressed as 
\begin{align}
{\hat{\bf{f}}}_{k_s} = {\bf{b}}_{k_s}\circ {\bf{q}}_{k_s},
\end{align}
and the channel error $\mathbf{e}_{k_{s}}$ follows $\mathbf{e}_{k_{s}} \sim \mathcal{CN}(\mathbf{0},{\sigma}_{e}^{2}\mathbf{I})$. 
% The vector $\mathbf{b}_{k_{s}}$ is composed of user terminal antenna gain, free space loss, and satellite beam radiation pattern. 
The $n_s$-th element of $\mathbf{b}_{k_{s}}$ is 
${b}_{k_s,n_s} = \frac{\sqrt{G_{R}G_{k_s,n_s}}}{4 \pi \frac{d_{k_s}}{\lambda}\sqrt{\kappa T_{sys} B_{w}}}$,
where $G_{R}$ is the user terminal antenna gain, $d_{k_{s}}$ is the distance between the satellite and SU-$k_{s}$, $\lambda$ is the carrier wavelength, $\kappa$ is the Boltzmann constant, $T_{sys}$ is the receiving system noise temperature, and $B_{w}$ is the bandwidth. 
In addition, the beam gain from the $n_{s}$-th antenna feed to SU-$k_{s}$ is expressed as ${G}_{k_s,n_s} = G_{max}\left[\frac{J_{1}(u_{k_s,n_s})}{2u_{k_s,n_s}} + 36\frac{J_{3}(u_{k_s,n_s})}{u_{k_s,n_s}^{3}}\right]^{2}$, where $u_{k_{s},n_{s}} = 2.07123 \mathrm{sin}(\theta_{k_{s},n_{s}})/\mathrm{sin}(\theta_{3\mathrm{dB}})$.
 $G_{max}$ denotes the maximum beam gain observed at each beam center.
 With the assumption that the position of each SU can be tracked, the angle between the center of beam-$n_{s}$ and SU-$k_{s}$ with respect to the satellite is denoted as $\theta_{k_{s},n_{s}}$ and 3dB loss angle compared with the beam center is $\theta_{3\mathrm{dB}}$. 
 Moreover, the rain attenuation effect and signal phase from the satellite and SU-$k_{s}$ are denoted by $\mathbf{q}_{k_{s}}$ of which the $n_{s}$-th element is ${q}_{k_s,n_s} = \chi_{k_s,n_s}^{-1/2}e^{-j\phi_{k_s,n_s}}$.
The dB form of rain attenuation gain is  $\chi_{k_{s},n_{s}}^{\rm{dB}} = 10 \mathrm{log}_{10}(\chi_{k_{s},n_{s}})$ following log-normal distribution with mean of $\mu$ and deviation of $\sigma$. The signal phase $\phi_{k_{s},n_{s}}$ is uniformly distributed between 0 and $2\pi$. 

Furthermore, the interfering channel matrix between the satellite and all CUs is expressed as $\mathbf{Z}=[\mathbf{z}_{1},...,\mathbf{z}_{K_{t}}] \in \mathbb{C}^{N_{s}\times K_{t}}$, where the vector $\mathbf{z}_{k_{t}}$ denotes channel between the satellite and CU-$k_{t}$. Likewise, the channel vector $\mathbf{z}_{k_{t}}$ is formulated as $\mathbf{z}_{k_{t}} = \hat{\mathbf{z}}_{k_{t}} + {\mathbf{e}}'_{k_{t}}$, where $\hat{\mathbf{z}}_{k_{t}}$ denotes channel estimate and ${\mathbf{e}}'_{k_{t}}$ denotes channel error following ${\mathbf{e}}'_{k_{t}} \sim \mathcal{CN}(\mathbf{0},{\sigma}_{e}^{2}\mathbf{I})$.

\begin{figure}[t]
    \centering
    %\vspace{-6mm}
    %\vspace{-5mm}
    \includegraphics[width=120mm]{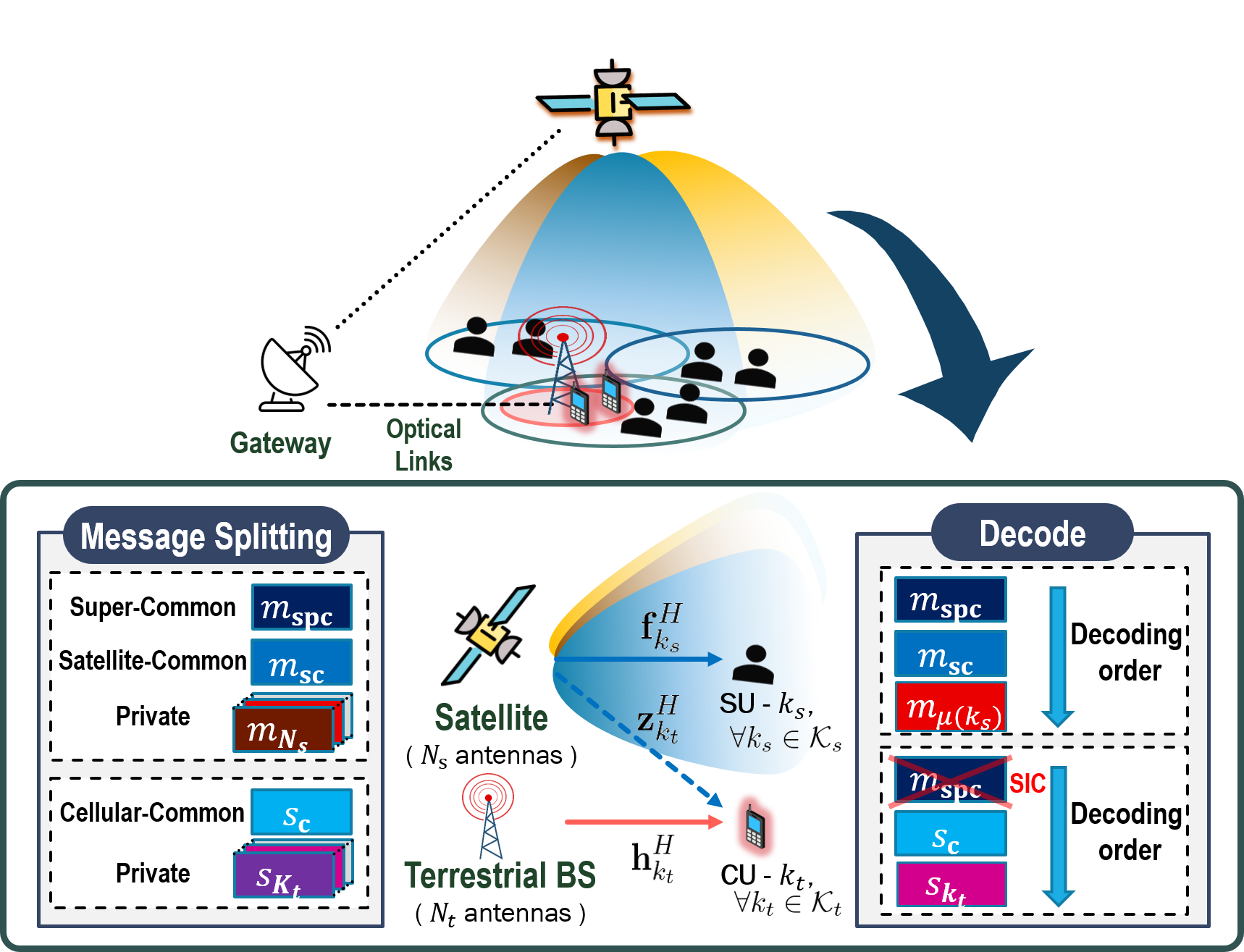}
    \caption{System architecture of coordinated RSMA with a super-common message for coordinated ISTN}
    %\vspace{-1mm}
    \label{fig:System Architecture}
    %\vspace{-5.5mm}
\end{figure}

\subsubsection{Terrestrial Channel Model}
The terrestrial channel matrix between the terrestrial BS and all CUs is expressed as $\mathbf{H}=[\mathbf{h}_{1},...,\mathbf{h}_{K_{t}}] \in \mathbb{C}^{N_{t}\times K_{t}}$, where $\mathbf{h}_{k_{t}}$ denotes channel vector between the BS and CU-$k_{t}$. We assume that the terrestrial channel follows Rayleigh fading channel model.
%\vspace{-5mm}
\subsection{Signal Model}
%\vspace{-0.5mm}
We propose a coordinated RSMA framework with the super-common message described in Fig.~\ref{fig:System Architecture}. 
The multicast messages of the satellite $M_{1},...,M_{N_{s}}$ intended for each beam are split into super-common parts, satellite-common parts, and private parts, i.e. $M_{n_{s}} \rightarrow \{M^{\sf{spc}}_{n_{s}}, M^{\sf{sc}}_{n_{s}}, M^{\sf{p}}_{n_{s}} \}$, $\forall n_{s} \in \mathcal{N}_{s}$. 
{All super-common parts are integrated into $M^{\sf{spc}}$ and encoded into super-common stream $m_{\sf{spc}}$ which can be decoded by all users through SIC, i.e. $\{M^{\sf{spc}}_{1},...,M^{\sf{spc}}_{N_{s}}\} \rightarrow m_{\sf{spc}}$.} 
%%It is assumed that this encoding process is operated according to a public codebook which is pre-shared with all SUs and CUs; thus, the $m_{\sf{spc}}$ can be decoded and deducted by all users via SIC. 
%When each user receives the signal, it initially decodes and recovers the $M^{\sf{spc}}$ via SIC through the predefined public codebook. 
Similarly, all satellite-common parts are combined as $M^{\sf{sc}}$ and encoded into satellite-common stream $m_{\sf{sc}}$, i.e. $\{M^{\sf{sc}}_{1},...,M^{\sf{sc}}_{N_{s}}\} \rightarrow m_{\sf{sc}}$. 
%%The encoding process of $m_{\sf{sc}}$ is also operated with its public codebook, which is pre-shared with all SUs. 
{The $m_{\sf{sc}}$ can be decoded and removed by all SUs through SIC after decoding super-common stream $m_{\sf{spc}}$.} 
%\hl{Similarly, the encoding and decoding process of the satellite-common message are operated with its pre-shared public codebook.} 
The private parts are encoded into each group's private stream independently, i.e. $M^{\sf{p}}_{n_{s}} \rightarrow m_{n_{s}}$. The vector of satellite streams can be formulated as $\mathbf{m}=[m_{\sf{spc}},m_{\sf{sc}},m_{1},...,m_{N_{s}}]^{T}\in\mathbb{C}^{(N_{s}+2)}$, where $\mathbb{E}\{\mathbf{m}\mathbf{m}^{H}\}=\mathbf{I}$. 
The beamforming matrix at the satellite is $\mathbf{W} = [\mathbf{w}_{\sf{spc}},\mathbf{w}_{\sf{sc}},\mathbf{w}_{1},...,\mathbf{w}_{N_{s}}]\in\mathbb{C}^{N_{s}\times(N_{s}+2)}$, where $\mathbf{w}_{\sf{spc}}$, $\mathbf{w}_{\sf{sc}}$, $\mathbf{w}_{n_{s}}$ are super-common precoder, satellite-common precoder, and private precoder for the $n_{s}$-th beam, respectively. 
%For the satellite, each antenna's per-feed power budget is given by $\{\mathbf{W}\mathbf{W}^{H}\}_{n_{s},n_{s}} \leq \frac{P_{s}}{N_{s}}, \forall n_{s} \in \mathcal{N}_{s}$.

{The unicast messages of the BS $W_{1},...,W_{K_{t}}$ intended for CUs are split into cellular-common and private parts, i.e. $W_{k_{t}} \rightarrow \{W_{\sf{c},\it{k_{t}}}, W_{\sf{p},\it{k_{t}}}\}, \forall k_{t}\in\mathcal{K}_{t}$. 
All cellular-common parts are combined as $W_{\sf{c}}$ and encoded into a cellular-common stream $s_{\sf{c}}$, i.e. $\{W_{\sf{c},1},...,W_{\sf{c},\it{K_{t}}}\} \rightarrow s_{\sf{c}}$, which can be decoded and removed by all CUs via SIC.} 
The private message for each CU is encoded into private stream independently, i.e. $W_{\sf{p}, \it{k_{t}}} \rightarrow s_{k_{t}}$. 
The vector of BS streams can be expressed as 
$\mathbf{s}=[s_{\sf{c}},s_{1},...,s_{K_{t}}]^{T}\in\mathbb{C}^{(K_{t}+1)}$, where $\mathbb{E}\{\mathbf{s}\mathbf{s}^{H}\}=\mathbf{I}$.
The beamforming matrix of BS is $\mathbf{P} = [\mathbf{p}_{\sf{c}},\mathbf{p}_{1},...,\mathbf{p}_{K_{t}}]\in\mathbb{C}^{{N_{t}} \times (K_{t}+1)}$, where $\mathbf{p}_{\sf{c}}$ and $\mathbf{p}_{k_{t}}$ are common precoder and CU-$k_{t}$'s private precoder, respectively.

{It is assumed that each encoding process for the common message is operated with its corresponding public codebook which is pre-shared with its intended users; thus, the common streams can be successfully decoded and removed by all intended users via SIC. 
The decoding process is performed first for the super-common stream regardless of its strength to perform the inter-network RS in advance of the intra-network RS. 
After the inter-network interference is partially decoded, the intra-network interference within each network is flexibly managed with the intra-network RS.
%The decoding process is operated in an order that data stream intended for more users has a higher priority of decoding, which is commonly used in the literature on RS \mbox{[\ref{ref:2018 EURASIP_Bruno_RSMA}]}, \mbox{[\ref{ref:HRS}]}, \mbox{[\ref{ref:RSMA_Survey}]}.
%That is, the common streams intended for more users are decoded and removed in priority, while the private streams are decoded after all of common streams are decoded and deducted.
}
Thus, the super-common stream is initially decoded and deducted through SIC, then the satellite-common/cellular-common streams and their desired private streams are decoded in order for all SUs/CUs.

The transmit signals at the satellite and BS are respectively
\begin{align}
\label{eq:Tx}
    \mathbf{x}^{\sf{sat}} = \mathbf{W}\mathbf{m}, \quad
    \mathbf{x}^{\sf{bs}} = \mathbf{P}\mathbf{s}.
\end{align}
The received signals at SU-$k_{s}$ and CU-$k_{t}$ can be written as
\begin{align}
\label{eq:Rx}
    \mathit{y}_{k_{s}}^{\sf{sat}} = \mathbf{f}_{k_{s}}^{H}\mathbf{W}\mathbf{m} + {n}_{k_{s}}^{\sf{sat}}, \!\!
    \quad
\mathit{y}_{k_{t}}^{\sf{bs}} = \mathbf{h}_{k_{t}}^{H} \mathbf{P}\mathbf{s} +\mathbf{z}_{k_{t}}^{H} \mathbf{W}\mathbf{m} + \mathit{n}_{k_{t}}^{\sf{bs}},
\end{align}
where $n_{k_{s}}^{\sf{sat}}\sim \mathcal{CN}(0,{\sigma_{k_{s}}^{\sf{sat} \rm{2}}})$ and $n_{k_{t}}^{\sf{bs}}\sim \mathcal{CN}(0,{\sigma_{k_{t}}^{\sf{bs} \rm{2}}})$ are the additive Gaussian white noises (AWGN) at SU-$k_{s}$ and CU-$k_{t}$, respectively.\footnote{Because the noise power is normalized by $\kappa T_{sys} B_{w}$ in the satellite channel model, the noise variances can be set as one, i.e., ${\sigma_{k_{s}}^{\sf{sat} \rm{2}}} = {\sigma_{k_{t}}^{\sf{bs} \rm{2}}} = 1$.} 
With limited knowledge of satellite channels, a concept of GMI is used to estimate the instantaneous achievable rates. 
%The achievable rates of super-common stream at SU-$k_{s}$ and CU-$k_{t}$ are respectively lower bounded as
The lower bounds of achievable rates of $m_{\sf{spc}}$ at SU-$k_{s}$ and CU-$k_{t}$ are respectively given by 
\begin{align}
\label{eq:GMI_Rate_spc_sat} 
    R_{\sf{spc}, \it{k_{s}}}^{\sf{sat}} \geq \mathrm{log}_{2}(1+\gamma_{\sf{spc}, \it{k_{s}}}^{\sf{sat}}), \!\!\!\quad
    R_{\sf{spc}, \it{k_{t}}}^{\sf{bs}} \geq \mathrm{log}_{2}(1+ \gamma_{\sf{spc}, \it{k_{t}}}^{\sf{bs}}),
    %\nonumber
\end{align}
in which $\gamma_{{\sf spc}, \it{k_{s}}}^{\sf{sat}} \triangleq \frac{|\hat{\mathbf{f}}_{k_{s}}^{H}\mathbf{w}_{\sf{spc}}|^{2}}{|\hat{\mathbf{f}}_{k_{s}}^{H}\mathbf{w}_{\sf{sc}}|^{2} + g}$, $\gamma_{\sf{spc}, \it{k_{t}}}^{\sf{bs}} \triangleq \frac{|\hat{\mathbf{z}}_{k_{t}}^{H} \mathbf{w}_{\sf{spc}}|^{2}} {|{\mathbf{h}}_{k_{t}}^{H} \mathbf{p}_{\sf{c}}|^{2} + l}$ is satisfied. 

Herein, $g \triangleq \sum_{i=1}^{N_{s}}|\hat{\mathbf{f}}_{k_{s}}^{H} \mathbf{w}_{i}|^{2} +\mathbb{E}\left[|\mathbf{e}_{k_{s}}^{H}\mathbf{w}_{\sf{spc}}|^{2} + |\mathbf{e}_{k_{s}}^{H}\mathbf{w}_{\sf{sc}}|^{2} + \sum_{i=1}^{N_{s}}|\mathbf{e}_{k_{s}}^{H}\mathbf{w}_{i}|^{2}\right] + \sigma_{k_{s}}^{\sf{sat} \rm{2}}$, and 
$l \triangleq \sum_{j=1}^{K_{t}}|\mathbf{h}_{k_{t}}^{H} \mathbf{p}_{j}|^{2} + |\hat{\mathbf{z}}_{k_{t}}^{H} \mathbf{w}_{\sf{sc}}|^{2}
+\sum_{i=1}^{N_{s}}|\hat{\mathbf{z}}_{k_{t}}^{H}\mathbf{w}_{i}|^{2} + \mathbb{E}\left[|\mathbf{e'}_{k_{t}}^{H}\mathbf{w}_{\sf{spc}}|^{2} + |\mathbf{e'}_{k_{t}}^{H}\mathbf{w}_{\sf{sc}}|^{2} + \sum_{i=1}^{N_{s}}|\mathbf{e'}_{k_{t}}^{H}\mathbf{w}_{i}|^{2}\right] + \sigma_{k_{t}}^{\sf{bs} \rm{2}}$ holds.

To guarantee that all users are able to decode $m_{\sf spc}$, the achievable rate of the $m_{\sf spc}$ is defined as
%\begin{align}
%\label{eq:sc-rate} 
    $R_{\sf{spc}} = \min \left( \!\min_{k_{s}\in \mathcal{K}_{s}}R_{\sf{spc},\it{k_{s}}}^{\sf{sat}}, \min_{k_{t}\in \mathcal{K}_{t}}R_{\sf{spc},\it{k_{t}}}^{\sf{bs}} \!\right)\! = \!\sum_{n_s=1}^{N_s}C^{\sf{spc}}_{n_s}$,
%\end{align}
where $C^{\sf{spc}}_{n_s}$ denotes the $n_{s}$-th beam's portion of super-common rate.
After $m_{\sf{spc}}$ is decoded and removed through SIC, all SUs (or CUs) decode the $m_{\sf{sc}}$ (or $s_{\sf{c}}$), and its corresponding private stream. 
{For the SU-$k_{s}$, the lower bounds of achievable rates of $m_{\sf sc}$ and $m_{\mu(k_{s})}$ are given by} 
%\begin{align}
%\label{eq:Rate_sc} 
    $R_{\sf{sc}, \it{k_{s}}}^{\sf{sat}} 
\geq \mathrm{log}_{2}(1+\gamma_{\sf{sc}, \it{k_{s}}}^{\sf{sat}}), \ 
R_{k_{s}}^{\sf{sat}} 
\geq \mathrm{log}_{2}(1+\gamma_{k_{s}}^{\sf{sat}})$,
%\end{align}
{respectively, in which $\gamma_{{\sf sc}, \it{k_{s}}}^{\sf{sat}} \triangleq \frac{|\hat{\mathbf{f}}_{k_{s}}^{H} \mathbf{w}_{\sf{sc}}|^{2}}{g}$, $\gamma_{k_{s}}^{\sf{sat}} \triangleq  \frac{|\hat{\mathbf{f}}_{k_{s}}^{H} \mathbf{w}_{\mu(k_{s})}|^{2}}
{g - |\hat{\mathbf{f}}_{k_{s}}^{H} \mathbf{w}_{\mu(k_{s})}|^{2}}$ is satisfied. Herein, $\mu(k_{s})$ denotes a mapping function between the SU-$k_{s}$ and its corresponding beam.} 
It corresponds to the satellite-common rate as $R_{\sf{sc}}^{\sf{sat}} = \min_{k_{s}\in \mathcal{K}_{s}}R_{\sf{sc},\it{k_{s}}}^{\sf{sat}}
= \sum_{n_s=1}^{N_s}C_{\sf{sc},\it{n_s}}^{\sf{sat}}$, where $C_{\sf{sc},\it{n_s}}^{\sf{sat}}$ is the $n_{s}$-th beam's portion of satellite-common rate.

{In addition, for the CU-$k_{t}$, the lower bounds of achievable rates of $s_{\sf c}$ and $s_{k_{t}}$ are respectively given by }
%\begin{align}
%\label{eq:Rate_common,private_CU} 
    $R_{\sf{c}, \it{k_{t}}}^{\sf{bs}} 
\geq \mathrm{log}_{2}(1+\gamma_{\sf{c}, \it{k_{t}}}^{\sf{bs}}), \ 
    R_{k_{t}}^{\sf{bs}} 
\geq \mathrm{log}_{2}(1+\gamma_{k_{t}}^{\sf{bs}})$,
%\end{align}
{where
%where the $\gamma_{\sf{c}, \it{k_{t}}}^{\sf{bs}}$ and $\gamma_{k_{t}}^{\sf{bs}}$ are respectively defined as
%\begin{align}
%\label{eq:SINR_common_CU}
    $\gamma_{{\sf c}, \it{k_{t}}}^{\sf{bs}} \triangleq \frac{|{\mathbf{h}}_{k_{t}}^{H} \mathbf{p}_{\sf{c}}|^{2}}
{l}, \gamma_{k_{t}}^{\sf{bs}} \triangleq \frac{|{\mathbf{h}}_{k_{t}}^{H} \mathbf{p}_{k_{t}}|^{2}}
{l - |{\mathbf{h}}_{k_{t}}^{H} \mathbf{p}_{k_{t}}|^{2}}$ is satisfied.}
It corresponds to the cellular-common rate as $R_{\sf{c}}^{\sf{bs}} 
= \min_{k_{t}\in \mathcal{K}_{t}}R_{c,k_{t}}^{\sf{bs}}
= \sum_{k_t=1}^{K_t}C_{k_t}^{\sf{bs}}$, where $C_{k_t}^{\sf{bs}}$ is the CU-$k_{t}$'s portion of cellular-common rate. 
Then, the total achievable rate for the $n_{s}$-th beam and CU-$k_{t}$ are formulated as
\begin{align}
%\label{eq:R_tot_sat} 
    R_{\sf{tot},\it{n_{s}}}^{\sf{sat}} = C^{\sf{spc}}_{\it{n_s}}+C_{\sf{sc},\it{n_{s}}}^{\sf{sat}} + \min_{i\in\mathcal{G}_{n_{s}}}R_{i}^{\sf{sat}}, \!\!\quad R_{\sf{tot},\it{k_{t}}}^{\sf{bs}} = C_{k_{t}}^{\sf{bs}} + R_{k_{t}}^{\sf{bs}},
    %\nonumber
\end{align}
where $\mathcal{G}_{n_{s}}$ denotes the set of SUs within the $n_{s}$-th beam.
% \vspace{-5mm}
\section{The Proposed Scheme}
% \vspace{-5mm}
%\subsection{\hl{Coordinated RS design for coordinated ISTN}}
{Our objective is to maximize the minimum rate of all users in the coordinated ISTN as the following problem $\mathcal{P}_{1}$. }
\begin{align} \label{eq:mmf_stage1}
\mathcal{P}_{1}:
&\mathop{{\text{max}}}_{{\bf{W}}, {\bf{P}}, {\bf{c}}^{\sf{spc}}, {\bf{c}}_{\sf{sc}}^{\sf{sat}}, {\bf{c}}^{\sf{bs}}} \mathop{\text{min}}_{n_{s} \in \mathcal{N}_{s}, k_{t} \in \mathcal{K}_{t}} \{R_{\sf{tot}, \it{k_{t}}}^{\sf{bs}}, R_{\sf{tot}, \it{n_{s}}}^{\sf{sat}}\}\\
{\text{s.t.}} 
&\;\;\;\;\;\; \mathit{R}_{\sf{spc}, \it{k_{s}}}^{\sf{sat}} \geq \sum_{j=1}^{N_{s}} C^{\sf{spc}}_{j}, \quad \mathit{R}_{\sf{spc}, \it{k_{t}}}^{\sf{bs}} \geq \sum_{j=1}^{N_{s}} C^{\sf{spc}}_{j},
\label{eq:constraint_1_mmf_stage1}  \\
%&\;\;\;\;\;\; C^{\sf{spc}}_{n_{s}} \geq 0,
%\label{eq:constraint_2_mmf_stage1} \\
&\;\;\;\;\;\; R_{\sf{sc}, \it{k_{s}}}^{\sf{sat}} \geq \sum_{j=1}^{N_{s}} C_{\sf{sc},\it{j}}^{\sf{sat}}, \quad R_{\sf{c}, \it{k_{t}}}^{\sf{bs}} \geq \sum_{j=1}^{K_{t}} C_{j}^{\sf{bs}},
\label{eq:constraint_3_mmf_stage1} \\
&\;\;\;\;\;\; C^{\sf{spc}}_{n_{s}} \geq 0, \quad C_{\sf{sc},\it{n_{s}}}^{\sf{sat}} \geq 0, \quad C_{k_{t}}^{\sf{bs}} \geq 0,
\label{eq:constraint_4_mmf_stage1} \\
&\;\;\;\;\;\; \mathrm{tr}\left(\mathbf{P P}^{H}\right) \leq P_{t}, \quad \left(\mathbf{W} \mathbf{W}^{H}\right)_{n_{s}, n_{s}} \leq \frac{P_{s}}{N_{s}},
\label{eq:constraint_5_mmf_stage1}
\end{align}
{where $\mathbf{c}^{\sf{spc}} = [C^{\sf{spc}}_{1},...,C^{\sf{spc}}_{N_{s}}]^{T}$,
$\mathbf{c}_{\sf{sc}}^{\sf{sat}} = [C_{1,\sf{sc}}^{\sf{sat}},...,C_{N_{s},\sf{sc}}^{\sf{sat}}]^{T}$,
$\mathbf{c}^{\sf{bs}} = [C_{1}^{\sf{bs}},...,C_{K_{t}}^{\sf{bs}}]^{T}$.
\mbox{\eqref{eq:constraint_1_mmf_stage1}} and \mbox{\eqref{eq:constraint_3_mmf_stage1}} ensure that $m_{\sf{spc}}$, $m_{\sf{sc}}$, and $s_{\sf{c}}$ can be decoded by their intended users. 
% \eqref{eq:constraint_3_mmf_stage1} guarantees $m_{\sf{sc}}$ and $s_{\sf{c}}$ to be decoded by all SUs/CUs. 
%\eqref{eq:constraint_3_mmf_stage1} guarantees $m_{\sf{sc}}$ to be decoded by all SUs, and $s_{\sf{c}}$ to be decoded by all CUs. 
\mbox{\eqref{eq:constraint_4_mmf_stage1}} represents that each portion of common rate is non-negative. 
%\eqref{eq:constraint_5_mmf_stage1} expresses the power budget at the BS and satellite. 
\mbox{\eqref{eq:constraint_5_mmf_stage1}} implies the power budget/per-feed power budget at BS/satellite. 
To convexify, we first rewrite $\mathcal{P}_{1}$ to apply SCA method \mbox{[\ref{ref:Arxiv_Bruno_RSMAinISTN}]} as 
}
% To convexify the problem $\mathcal{P}_{1}$, the SCA-based method in \mbox{[\ref{ref:Arxiv_Bruno_RSMAinISTN}]} is employed.
% First, we reformulate the $\mathcal{P}_{1}$ equivalently as 
% \vspace{-2mm}  
% To build the convex version of $\mathcal{P}_1$, we stepwise reformulate $\mathcal{P}_1$ as $\mathcal{P}_2$ and $\mathcal{P}_3$.
% To convexify the non-convex rate constraints, the problem can be rewritten
\begin{align} \label{eq:mmf_stage2}
\mathcal{P}_{2}:
&\mathop{{\text{max}}}_{{\bf{W}}, {\bf{P}}, {\bf{c}}^{\sf{spc}}, {\bf{c}}_{\sf{sc}}^{\sf{sat}}, {\bf{c}}^{\sf{bs}}, q, \bf{\alpha}, \bf{r}} q\\
 {\text{s.t.}} \quad
&C^{\sf{spc}}_ {\it{n_{s}}} + C_{\sf{sc}, \it{n_{s}}}^{\sf{sat}} + r_{k_{s}} \geq q, \quad 
\forall k_{s} \in \mathcal{G}_{n_{s}},
\label{eq:constraint_1_mmf_stage2}  \\
&C_{k_{t}}^{\sf{bs}} + \alpha_{k_{t}} \geq q, \quad \forall k_{t} \in \mathcal{K}_{t},
\label{eq:constraint_2_mmf_stage2}  \\
&R_{k_{s}}^{\sf{sat}} \geq r_{k_{s}}, \ R_{k_{t}}^{\sf{bs}} \geq \alpha_{k_{t}}, \ \forall k_{s} \in \mathcal{K}_{s}, \ \forall k_{t} \in \mathcal{K}_{t},
\label{eq:constraint_3_mmf_stage2}  \\
%&R_{k_{t}}^{\sf{bs}} \geq \alpha_{k_{t}}, \quad \forall k_{t} \in \mathcal{K}_{t},
%\label{eq:constraint_4_mmf_stage2}  \\
& \eqref{eq:constraint_1_mmf_stage1}  - \eqref{eq:constraint_5_mmf_stage1}, \nonumber
\end{align}
{where $q$, $\mathbf{\alpha} = \left[\alpha_{1},...,\alpha_{K_{t}}\right]^{T}$, and $\mathbf{r} = \left[r_{1},...,r_{K_{s}}\right]^{T}$ are some auxiliary variables. 
Furthermore, it can be rewritten by}
\begin{align} \label{eq:mmf_stage3}
\mathcal{P}_{3}:
&\mathop{{\text{max}}}_{{\bf{W}}, {\bf{P}}, {\bf{c}}_{\sf{spc}}, {\bf{c}}_{\sf{sc}}^{\sf{sat}}, {\bf{c}}^{\sf{bs}}, q, \bf{\alpha}, \bf{r}, \bf{a}, \bf{a}_{\sf{spc}}, \bf{a}_{\sf{c}}, \bf{b}, \bf{b}_{\sf{spc}}, \bf{b}_{\sf{sc}}}  q \\
{\text{s.t.}} \quad
&\mathrm{ln}(1+a_{\sf{spc},\it{k_{t}}}) \geq \sum_{j=1}^{N_{s}}C^{\sf{spc}}_{\it{j}} \mathrm{ln}2, 
\label{eq:constraint_1_mmf_stage3}  \\
&\gamma_{\sf{spc},\it{k_{t}}}^{\sf{bs}} \geq a_{\sf{spc},\it{k_{t}}},
\label{eq:constraint_2_mmf_stage3}  \\
&\mathrm{ln}(1+b_{\sf{spc},\it{k_{s}}}) \geq \sum_{j=1}^{N_{s}}C^{\sf{spc}}_{\it{j}}\mathrm{ln}2,
\label{eq:constraint_3_mmf_stage3}  \\
&\gamma_{\sf{spc},\it{k_{s}}}^{\sf{sat}} \geq b_{\sf{spc},\it{k_{s}}},
\label{eq:constraint_4_mmf_stage3}  \\
&\mathrm{ln}(1+a_{\sf{c},\it{k_{t}}}) \geq \sum_{j=1}^{K_{t}}C_{j}^{\sf{bs}} \mathrm{ln}2, \ \gamma_{\sf{c},\it{k_{t}}}^{\sf{bs}} \geq a_{\sf{c},\it{k_{t}}},
\label{eq:constraint_5_mmf_stage3}  \\
&\mathrm{ln}(1+b_{{\sf sc},k_{s}}) \geq \sum_{j=1}^{N_{s}}C_{{\sf sc},j}^{\sf{sat}} \mathrm{ln}2, \quad \gamma_{\sf{sc},\it{k_{s}}}^{\sf{sat}} \geq b_{\sf{sc},\it{k_{s}}}, 
\label{eq:constraint_6_mmf_stage3} \\
&\mathrm{ln}(1+a_{k_{t}}) \geq \alpha_{k_{t}} \mathrm{ln}2, \quad \gamma_{k_{t}}^{\sf{bs}} \geq a_{k_{t}},
\label{eq:constraint_7_mmf_stage3}  \\
&\mathrm{ln}(1+b_{k_{s}}) \geq r_{k_{s}} \mathrm{ln}2, \quad \gamma_{k_{s}}^{\sf{sat}} \geq b_{k_{s}},
\label{eq:constraint_8_mmf_stage3}  \\
&\eqref{eq:constraint_4_mmf_stage1}, \ \eqref{eq:constraint_5_mmf_stage1}, \ \eqref{eq:constraint_1_mmf_stage2}, \  \eqref{eq:constraint_2_mmf_stage2}, \nonumber
\end{align}
for all $k_{s}\in\mathcal{K}_{s}$ and $k_{t}\in\mathcal{K}_{t}$, 
where $\mathbf{a} = \left[a_{1},...,a_{K_{t}}\right]^{T}$,
$\mathbf{a}_{\sf{spc}} = \left[a_{\sf{spc},1},...,a_{\sf{spc},\it{K_{t}}}\right]^{T}$,
$\mathbf{a}_{\sf{c}} = \left[a_{\sf{c},1},...,a_{\sf{c},\it{K_{t}}}\right]^{T}$,
$\mathbf{b} = \left[b_{1},...,b_{K_{s}}\right]^{T}$,
$\mathbf{b}_{\sf{spc}} = \left[b_{\sf{spc},1},...,b_{\sf{spc},\it{K_{s}}}\right]^{T}$,
$\mathbf{b}_{\sf{sc}} = \left[b_{\sf{sc},1},...,b_{\sf{sc},\it{K_{s}}}\right]^{T}$
are further introduced auxiliary variables.
The constraint \eqref{eq:constraint_2_mmf_stage3} can be expanded by 
\begin{align}
\label{eq:constraint_2-1_mmf_stage3}
|{\mathbf{h}}_{k_{t}}^{H} \mathbf{p}_{\sf{c}}|^{2} + l \leq  \frac{|\hat{\mathbf{z}}_{k_{t}}^{H} \mathbf{w}_{\sf{spc}}|^{2}}{a_{\sf{spc},\it{k_{t}}}}.
\end{align}
{To use the SCA method, the RHS term $ \frac{|\hat{\mathbf{z}}_{k_{t}}^{H} \mathbf{w}_{\sf{spc}}|^{2}}{a_{\sf{spc},\it{k_{t}}}}$ can be approximated and lower-bounded by first-order Taylor series expansion around the point $(\mathbf{w}_{\sf{spc}}^{[n]},a_{\sf{spc},\it{k_{t}}}^{[n]})$ as}
\begin{align}
    \frac{|\hat{\mathbf{z}}_{k_{t}}^{H} \mathbf{w}_{\sf{spc}}|^{2}}{a_{\sf{spc},\it{k_{t}}}} &\geq
\frac{2\mathrm{Re}[\mathbf{w}_{\sf{spc}}^{[n]H} \hat{\mathbf{z}}_{k_{t}} \hat{\mathbf{z}}_{k_{t}}^{H} \mathbf{w}_{\sf{spc}}]}{a_{\sf{spc},\it{k_{t}}}^{[n]}} - \frac{\mathbf{w}_{\sf{spc}}^{[n]H} \hat{\mathbf{z}}_{k_{t}} \hat{\mathbf{z}}_{k_{t}}^{H} \mathbf{w}_{\sf{spc}}^{[n]}}{(a_{\sf{spc},\it{k_{t}}}^{[n]})^{2}}a_{\sf{spc},\it{k_{t}}} \nonumber\\  
&\triangleq \hat{f_{1}}(\mathbf{w}_{\sf{spc}},a_{\sf{spc},\it{k_{t}}};\mathbf{w}_{\sf{spc}}^{\mathit{[n]}},a_{\sf{spc},\it{k_{t}}}^{\mathit{[n]}}),
\end{align}
{where the $n$ denotes the $n$-th SCA iteration.} 
Then, the constraint \mbox{\eqref{eq:constraint_2-1_mmf_stage3}} can be near equivalently approximated by 
\begin{align}
\label{eq:constraint_2-2_mmf_stage3}
|{\mathbf{h}}_{k_{t}}^{H} \mathbf{p}_{\sf{c}}|^{2} + l \leq  \hat{f_{1}}(\mathbf{w}_{\sf{spc}},a_{\sf{spc},\it{k_{t}}};\mathbf{w}_{\sf{spc}}^{\mathit{[n]}},a_{\sf{spc},\it{k_{t}}}^{\mathit{[n]}}),
\end{align}
which is a convex form. 
Likewise, the constraint \mbox{\eqref{eq:constraint_4_mmf_stage3}} can be convexified as
\begin{align}
\label{eq:constraint_4-2_mmf_stage3}
|\hat{\mathbf{f}}_{k_{s}}^{H} \mathbf{w}_{\sf{sc}}|^{2} + g \leq \hat{f_{2}}(\mathbf{w}_{\sf{spc}},b_{\sf{spc},\it{k_{s}}};\mathbf{w}_{\sf{spc}}^{\mathit{[n]}},b_{k_{s}}^{\mathit{[n]}}),
\end{align}
where $\hat{f_{2}}(\mathbf{w}_{\sf{spc}},b_{\sf{spc},\it{k_{s}}};\mathbf{w}_{\sf{spc}}^{\mathit{[n]}},b_{\sf{spc},\it{k_{s}}}^{\mathit{[n]}}) \triangleq \frac{2\mathrm{Re}[\mathbf{w}_{\sf{spc}}^{[n]H} {\hat{\mathbf{f}}}_{k_{s}} {\hat{\mathbf{f}}}_{k_{s}}^{H} \mathbf{w}_{\sf{spc}}]}{b_{\sf{spc},\it{k_{s}}}^{[n]}} - \frac{\mathbf{w}_{\sf{spc}}^{[n]H} {\hat{\mathbf{f}}}_{k_{s}} {\hat{\mathbf{f}}}_{k_{s}}^{H} \mathbf{w}_{\sf{spc}}^{[n]}}{(b_{\sf{spc},\it{k_{s}}}^{[n]})^{2}}b_{\sf{spc},\it{k_{s}}}$ holds. 
In the constraint of \mbox{\eqref{eq:constraint_1_mmf_stage3}}, as the log term is a general non-linear convex program, it can lead to the high overhead of per-iteration computation in the CVX toolbox. To alleviate the computational complexity, we equivalently reformulate it into a second-order cone (SOC) constraint as described in \mbox{[\ref{ref:Arxiv_Bruno_RSMAinISTN}]}, which is \begin{align}
 ||[a_{\sf{spc},\it{k_{t}}} + \sum_{j=1}^{N_{s}}C^{{\sf spc}}_{j}\mathrm{log}2 - {v}_{\sf{spc},\it{k_{t}}}^{[n]} \quad 2\sqrt{{u}_{\sf{spc},\it{k_{t}}}^{[n]}}]||_{2} 
 \leq a_{\sf{spc},\it{k_{t}}} -  \sum_{j=1}^{N_{s}}C^{\sf{spc}}_{j}\mathrm{log}2 + {v}_{\sf{spc},\it{k_{t}}}^{[n]},
\end{align}
where ${v}_{\sf{spc},\it{k_{t}}}^{[n]} = \frac{a_{\sf{spc},\it{k_{t}}}^{[n]}}{a_{\sf{spc},\it{k_{t}}}^{[n]} + 1} + \mathrm{ln}(a_{\sf{spc},\it{k_{t}}}^{[n]} + 1)$, ${u}_{\sf{spc},\it{k_{t}}}^{[n]} = \frac{(a_{\sf{spc},\it{k_{t}}}^{[n]})^{2}}{a_{\sf{spc},\it{k_{t}}}^{[n]} + 1}$ holds.
Similarly, the constraint \eqref{eq:constraint_3_mmf_stage3} can be reformulated by
%\vspace{-2mm}
\begin{align}
||[b_{{\sf spc}   ,k_{s}} + \sum_{j=1}^{N_{s}}C^{{\sf spc}}_{j}\mathrm{log}2 - \bar{v}_{{\sf spc},k_{s}}^{[n]} \quad 2\sqrt{\bar{u}_{{\sf spc},k_{s}}^{[n]}}]||_{2} 
\leq b_{{\sf spc},k_{s}} -  \sum_{j=1}^{N_{s}}C^{{\sf spc}}_{j}\mathrm{log}2 + \bar{v}_{{\sf spc},k_{s}}^{[n]},
\end{align}
where $\bar{v}_{{\sf spc},k_{s}}^{[n]} = \frac{b_{{\sf spc},k_{s}}^{[n]}}{b_{{\sf spc},k_{s}}^{[n]} + 1} + \mathrm{ln}(b_{{\sf spc},k_{s}}^{[n]} + 1)$, $\bar{u}_{{\sf spc},k_{s}}^{[n]} = \frac{(b_{{\sf spc},k_{s}}^{[n]})^{2}}{b_{{\sf spc},k_{s}}^{[n]} + 1}$.
Other constraints \mbox{\eqref{eq:constraint_5_mmf_stage3}} - \mbox{\eqref{eq:constraint_8_mmf_stage3}} also can be approximated into convex and SoC constraints as described in \mbox{[\ref{ref:Arxiv_Bruno_RSMAinISTN}]}, then the problem $\mathcal{P}_{3}$ can be reformulated into convex and computational efficient problem formulation.\footnote{
As the algorithm is iterated until stopping criterion $|q^{[n]}-q^{[n-1]}|<10^{-4}$, the objective function value $q$ monotonically increases and it is bounded above by the power budgets. Although the solution of our SCA-based algorithm is sub-optimal, it converges to the set of stationary points of problem $\mathcal{P}_{1}$ \mbox{[\ref{ref:Math_Convex}]}. 
The studies for globally optimal beamforming for RSMA-based ISTN and the comparison of the results with sub-optimal solutions in this paper are left for future work.}

%\hl{Remark (Difference with hierarchical rate splitting} [\ref{ref:HRS}]\hl{): 
%{\bf{Remark (Difference with hierarchical rate splitting [ref:HRS])}}:
 {\bf Remark 1 (Difference with hierarchical rate splitting} \mbox{[\ref{ref:HRS}]}
  {\bf ):}
 %{\bf \hl{Remark 1 (Difference with hierarchical rate splitting \mbox{[\ref{ref:HRS}]}
 %):}}
 \it{
In hierarchical rate splitting (HRS), there are private messages intended for each user and two kinds of common messages, namely an outer-tier common message and an inner-tier common message, that can be decoded by all users and by a subset of users, respectively. 
    Our proposed framework is similar to the HRS scheme in the sense that the two-layer SIC is needed to decode the different levels of the common messages. 
    Nevertheless, the proposed framework fundamentally differs from HRS because the transmitted data information for all users is shared at the transmitter in the HRS (particularly suited to multi-antenna broadcast channels), while it is not exchanged between the satellite and terrestrial BS in our system model. 
    In addition, as we consider the multibeam multicast transmission scenario in the satellite network, the intra-beam interference does not exist in our scenario, while it is considered in the HRS and managed by the inner-tier common message. 
    For these reasons, the existing HRS is not directly applicable to our scenario.
}
%\vspace{-3mm}
\section{Numerical Results}
\begin{figure}[!t]
    \centering
  \vspace{-80mm}   
  \includegraphics[width=170mm]{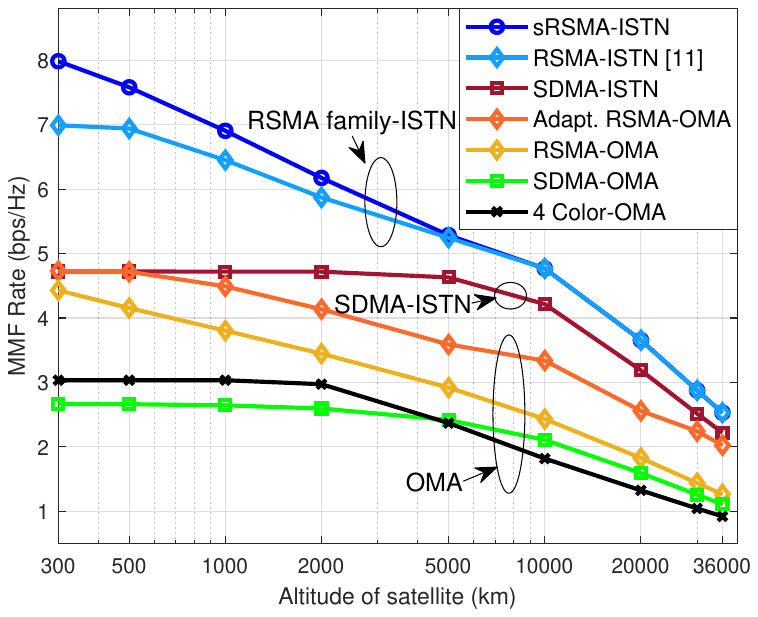}
 \vspace{-80mm}
 % \vspace{-3mm}
  %\vspace{-5mm}
    \caption{{MMF rate versus $h_{\sf{sat}}$, $P_{t} = 30\;\mathrm{dBm}$}}
    %\vspace{-4mm}
    \label{fig:Altitude}
\end{figure}

\begin{figure}
    \centering   
      \vspace{-80mm}   
  \includegraphics[width=170mm]{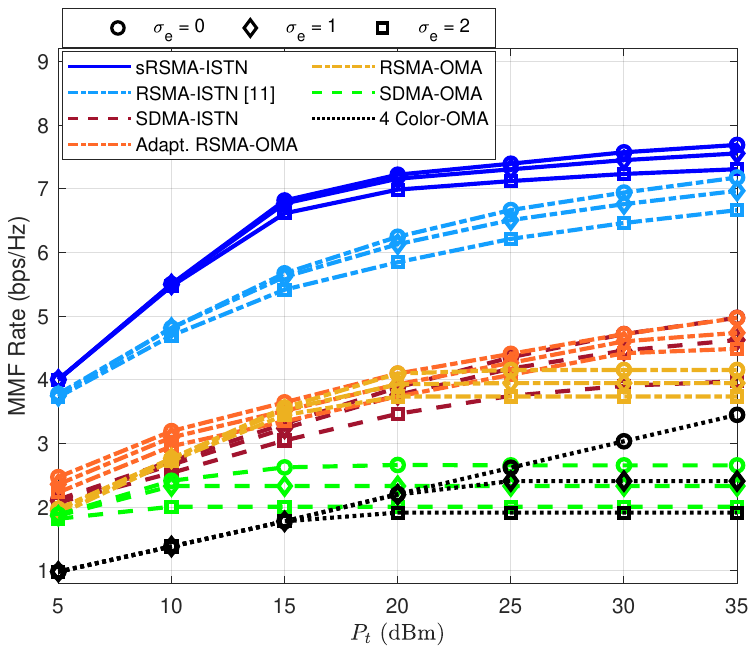}
   % \vspace{-3mm}
    %\vspace{-5mm}
    \vspace{-80mm}
    \caption{MMF rate versus $P_{t}$, $h_{\sf{sat}} = 500\;\mathrm{km}$} 
\label{fig:ImperfectCSI}
\end{figure}

\rm{In this section, the performance of our proposed framework is evaluated through numerical simulation, compared with conventional RSMA, SDMA, OMA schemes.} 
The simulation parameters are provided in Table \uppercase\expandafter{\romannumeral1}.
In Fig.~\ref{fig:Altitude} and Fig.~\ref{fig:ImperfectCSI}, the label ``$\sf{sRSMA\mbox{-}ISTN}$'' refers to coordinated RSMA with the super-common message in ISTN (our proposed); the labels ``$\sf{RSMA\mbox{-}ISTN}$'' and ``$\sf{SDMA\mbox{-}ISTN}$'' refer to the conventional RSMA without super-common message and SDMA are employed in ISTN, respectively. 
The label ``$\sf{Adapt.}$ $\sf{RSMA\mbox{-}OMA}$'' refers to an adaptive orthogonal multiple access (OMA) where the frequency band is dynamically divided into two parts (one for the satellite network and the other for the cellular network), and RSMA is separately employed in each network so as to maximize the MMF rate. 
The label ``$\sf{RSMA\mbox{-}OMA}$'' and ``$\sf{SDMA\mbox{-}OMA}$'' refer to schemes in which RSMA and SDMA are used with OMA scheme, respectively.
``$\sf{4}$ $\sf{Color\mbox{-}OMA}$'' refers to an OMA scheme where  the frequency band for each beam is reused with four orthogonal bands.

\begin {table}[t]
\caption {Simulation Parameters} 
%\vspace{-3.0mm}
\label{parameter} 
\begin{center}
\begin{tabular}{ c|c|c }
 \hlineB{3}
 {\textbf{Abbreviation}} & {\textbf{Definition}} & {\textbf{Value} } \\
 \hline
 %\hline\hline
 $f_{\rm{c}}$ & Frequency band (carrier frequency) & Ka (28 GHz) \\
 %\hline
 $B_{\it{w}}$ & Bandwidth & 500 MHz \\ 
 %\hline
 $\theta_{\rm{3dB}}$ & 3 dB angle & $0.4^{\circ}$ \\
 %\hline
 $G_{max}$ & Maximum beam gain & 52 dBi \\
 %\hline
 $G_{R}$ & User terminal antenna gain & 42.7 dBi \\
 %\hline
 $(\mu,\sigma)$ & Rain fading parameters & (-3.125,1.591) \\
 %\hline
 $N_{s}$ & Number of antennas at satellite & 3 \\
 %\hline
 $\rho$ & Number of SUs in each beam & 2 \\
 %\hline
 $N_{t}$ & Number of antennas at BS & 16 \\
 %\hline
 $K_{t}$ & Total number of CUs & 3 \\
 %\hline
 $P_{s}$ & Power budget at satellite & $50$ $\mathrm{W}$ \\
 \hlineB{3}
\end{tabular}
\end{center}
%\vspace{-6mm}
\end {table}

At first, we evaluate the MMF performance of our proposed framework versus the altitude of the satellite $h_{\sf sat}$ with perfect CSI assumption at the GW, i.e. $\sigma_e=0$.
The power budget at the BS $P_{t}$ is fixed at 30 dBm.
In Fig.~\ref{fig:Altitude}, as the altitude gets higher, the MMF rates of all schemes decrease with the degradation of satellite channel strength.
{With partially decoding intra-network (within satellite network and cellular network) interference and partially treating it as noise, all of the RSMA schemes outperform SDMA schemes in all cases of frequency band usage (ISTN and OMA); moreover, the $\sf{sRSMA\mbox{-}ISTN}$ has a further improved MMF performance by flexible management of inter-network interference with the super-common message. 
%performance of $\sf{sRSMA\mbox{-}ISTN}$ is further improved by flexible management of inter-network interference with the super-common message. 
In general, the performance gap between $\sf{sRSMA\mbox{-}ISTN}$ and $\sf{RSMA\mbox{-}ISTN}$ is enhanced by decreasing $h_{\sf{sat}}$; it becomes substantial particularly in the LEO region due to the high level of inter-network interference. 
%It has a substantial performance gap to $\sf{RSMA\mbox{-}ISTN}$ especially in the LEO region due to high inter-network interference level. 
%In general, the performance gap between $\sf{sRSMA\mbox{-}ISTN}$ and $\sf{RSMA\mbox{-}ISTN}$ is enhanced by lowering $h_{\sf{sat}}$; it becomes substantial especially in the LEO region due to the high level of inter-network interference. 
Especially when the satellite is at the very low Earth orbit (VLEO) with an altitude of $h_{\sf sat}=300$ km, the MMF rate of $\sf{sRSMA\mbox{-}ISTN}$ is increased by 14.2 \% than that of $\sf{RSMA\mbox{-}ISTN}$. 
The high level of inter-network interference in the LEO region can be mitigated by assigning lots of power to the super-common message, while it is fully treated as noise in $\sf{RSMA\mbox{-}ISTN}$.\footnote{{%Although the complexity in our proposed scheme is higher than in the scheme of \mbox{[\ref{ref:Arxiv_Bruno_RSMAinISTN}]}, it has an improved MMF rate, especially in the LEO region. 
Our proposed scheme achieves a significantly improved MMF rate in the LEO regime at the cost of a bit higher computational complexity than $\sf{RSMA\mbox{-}ISTN}$  \mbox{[\ref{ref:Arxiv_Bruno_RSMAinISTN}]} induced by an additional layer of SIC at the receivers.
%requires a bit higher complexity than the scheme in  due to the addition of one layer of SIC, it has an improved MMF rate, especially in the LEO region. 
%This implies that our proposed scheme can be practically beneficial given the low latency and launch costs of LEO satellites.
%The studies for the tradeoff between MMF rate and complexity are left for future work.
}}
On the other hand, as the level of inter-network interference becomes low by increasing the $h_{\sf sat}$, the power allocated to the super-common message is reduced to almost zero when the $h_{\sf sat}$ gets higher than 10,000 km. 
This makes the inter-network interference be fully treated as noise, causing the  $\sf{sRSMA\mbox{-}ISTN}$ scheme to have MMF performance similar to the ${\sf RSMA\mbox{-}ISTN}$ scheme in the region with a higher altitude than 10,000 km.
In OMA schemes, the frequency spectrum cannot be used more efficiently than in ISTN schemes with full frequency reuse, resulting in much worse performance.} 
In case of $\sf{Adapt.}$ $\sf{RSMA\mbox{-}OMA}$, although the frequency band is optimally divided between satellite/cellular networks with the management of the inter-beam/intra-cell interference, it has a lower performance than the ISTN-based schemes. 
For the scheme of $\sf{4}$ $\sf{Color\mbox{-}OMA}$, since the inter-beam interference is alleviated, it can have a higher performance than $\sf{SDMA\mbox{-}OMA}$ at LEO region. 
As the altitude of satellite $h_{\sf{sat}}$ grows, it gets the worst performance among the schemes, since the minimum rate is obtained by the rates of SUs which are provided with much narrower frequency bands than other schemes. 
Furthermore, in Fig.~\ref{fig:ImperfectCSI}, we evaluate the MMF performance versus terrestrial $P_{t}$ with a fixed altitude of satellite $h_{\sf sat} = 500$ $\rm{km}$, assuming imperfect satellite CSI at the GW.
Fig.~\mbox{\ref{fig:ImperfectCSI}} also shows the superiority of $\sf{sRSMA\mbox{-}ISTN}$ over other baseline schemes.
The MMF rates of overall schemes increase gradually as $P_{t}$ increases, yet they are saturated in the high $P_{t}$ region due to the fixed satellite power budget $P_{s}$. 
The intuition behind this is that in the high $P_{t}$ region, the minimum rate of all users equals the minimum rate of SUs, causing saturated performance due to the fixed satellite power budget $P_{s}$. On the other hand, for the case of perfect CSI (i.e. $\sigma_e=0$) in the scheme of $\sf{4}$ $\sf{Color\mbox{-}OMA}$, since the inter-beam interference is relieved, the MMF rate is determined by the minimum rate of CUs, thus it increases proportionally with $P_{t}$. 
We can also observe performance degradation in all schemes with increasing satellite CSI error levels. 
Specifically, the performance gaps among RSMA-based schemes with different levels of CSI errors are narrower than other schemes; the performance degradation of $\sf{sRSMA\mbox{-}ISTN}$ is much smaller than that of $\sf{RSMA\mbox{-}ISTN}$ in the presence of CSI errors. 
{These decreased performance gaps show that RSMA is more robust to CSI error than SDMA and 4 Color schemes, and our proposed framework has far better robustness through the flexible management of inter-network interference by partially decoding it and partially treating it as noise.}
%\vspace{-0.5mm} 
%\vspace{-4.3mm}
%\vspace{-3mm}
\section{Conclusion} 
%\vspace{-1mm}
In this paper, a coordinated RSMA framework with the super-common message for a coordinated ISTN  has been investigated. 
With imperfect satellite CSI at the GW, the joint beamforming optimization problem with power budgets has been tackled, aiming to achieve the MMF among all users in the coordinated ISTN.
The presented simulation results have numerically shown the superiority and satellite CSI-error robustness of our proposed framework. 
Thus, our proposed framework in ISTN is envisioned to fulfill ubiquitous global connectivity for  next-generation wireless networks.

%\section*{Acknowledgment}


\ifCLASSOPTIONcaptionsoff
  \newpage
\fi

%\vspace{-4.7mm}
%\vspace{-3mm}

%It is not necessary to upload the biography when you submit your manuscript.


\begin{thebibliography}{1}
%\vspace{-0.5mm}
% You can use other form of bib file by changing here...
%\bibitem{IEEEhowto:kopka} %\label{ref:TVT_Multicast beamforming optimization}
%H. Zhang, C. Jiang, J. Wang, L. Wang, Y. Ren, and L. Hanzo, "Multicast beamforming optimization in cloud-based heterogeneous terrestrial and satellite networks," \emph{IEEE Trans. Veh. Technol.}, vol. 69, no. 2, pp. 1766-1776, 2019.
\bibitem{IEEEhowto:kopka} \label{ref:Access_Survey}
P. Wang, J. Zhang, X. Zhang, Z. Yan, B. G. Evans, and W. Wang,  ``Convergence of Satellite and Terrestrial Networks: A Comprehensive Survey,''  \emph{IEEE Access}, vol. 8, pp. 5550-5588, Dec. 2019.
\bibitem{IEEEhowto:kopka} \label{ref:Magazine_Signal Processing for HTS}
A. I. Perez-Neira, M. A. Vazquez, M. R. B. Shankar, S. Maleki, and
S. Chatzinotas, ``Signal Processing for High-Throughput Satellites:
Challenges in New Interference-Limited Scenarios,''  \emph{IEEE Signal Process. Mag.}, vol. 36, no. 4, pp. 112-131, Jul. 2019.
%\bibitem{IEEEhowto:kopka} \label{ref:JSAC_cooperative multicast in ISTN}
%X. Zhu, C. Jiang, L. Yin, N. Ge, and J. Lu, ``Cooperative multigroup multicast transmission in integrated terrestrial-satellite networks,''  \emph{IEEE J. Sel. Areas Commun.}, vol. 36, no. 5, pp. 981-992, 2018.
\bibitem{IEEEhowto:kopka} \label{ref:2019 Letter_Bruno_RSMA}
B. Clerckx, Y. Mao, R. Schober, and H. V. Poor, ``Rate-Splitting Unifying SDMA, OMA, NOMA, and Multicasting in MISO Broadcast Channel: A Simple Two-User Rate Analysis,''  \emph{IEEE Wireless Commun. Lett.}, vol. 9, no. 3, pp. 349-353, Nov. 2019.
\bibitem{IEEEhowto:kopka} \label{ref:2018 EURASIP_Bruno_RSMA}
Y. Mao, B. Clerckx, and V. O. K. Li, ``Rate-splitting multiple access for downlink communication systems: bridging,
generalizing, and outperforming SDMA and NOMA,'' \emph{EURASIP J. Wireless Commun. Netw.}, no. 1, p. 133, May 2018.
\bibitem{IEEEhowto:kopka} \label{ref:2016 TCOM_Bruno_RSMA imperfect CSIT}
H. Joudeh and B. Clerckx, ``Sum-Rate Maximization for Linearly Precoded Downlink Multiuser MISO Systems With Partial CSIT: A Rate-Splitting Approach,''  \emph{IEEE Trans. Commun.}, vol. 64, no. 11, pp. 4847-4861, Aug. 2016.
\bibitem{IEEEhowto:kopka} \label{ref:2015 TCOM_Bruno_RSMA imperfect CSIT}
C. Hao, Y. Wu, and B. Clerckx, ``Rate Analysis of Two-Receiver MISO Broadcast Channel With Finite Rate Feedback: A Rate-Splitting Approach,''  \emph{IEEE Trans. Commun.}, vol. 63, no. 9, pp. 3232-3246, Jul. 2015.
\bibitem{IEEEhowto:kopka} \label{ref:2016 TSP_Bruno_RSMA imperfect CSIT}
H. Joudeh and B. Clerckx, ``Robust Transmission in Downlink Multiuser MISO Systems: A Rate-Splitting Approach,''  \emph{IEEE Trans. Signal Process.}, vol. 64, no. 23, pp. 6227-6242, Dec. 2016.
%\label{ref:2018 ISWCS_RSMA for Multibeam Satellite}
%M. Caus, A. Pastore, M. Navarro, T. Ramirez, C. Mosquera, N. Noels, N. Alagha, and A. I. Perez-Neria, ``Exploratory Analysis of Superposition Coding and Rate Splitting for Multibeam Satellite Systems,'' in \emph{Proc. IEEE Int. Symp. Wireless Commun. Syst. (ISWCS)}, Aug. 2018, pp. 1-5.
%\bibitem{IEEEhowto:kopka} \label{ref:ICC_RSMA for Multibeam Satellite}
%L. Yin and B. Clerckx, "Rate-splitting multiple access for multibeam satellite communications," in \emph{Proc. IEEE Int. Conf. Commun.(ICC) Workshop}, 2020, pp. 1-6.
\bibitem{IEEEhowto:kopka} \label{ref:TCOM_RSMA for Multibeam Satellite}
L. Yin and B. Clerckx, ``Rate-Splitting Multiple Access for Multigroup Multicast and Multibeam Satellite Systems,''  \emph{IEEE Trans. Commun.}, vol. 69, no. 2, pp. 976-990, Nov. 2020.
\bibitem{IEEEhowto:kopka} \label{ref:2021ICC_Bruno_RSMAforSatellite}
L. Yin, O. Dizdar, and B. Clerckx, ``Rate-Splitting Multiple Access for Multigroup Multicast Cellular and Satellite Communications: PHY Layer Design and Link-Level Simulations,'' in \emph{Proc. IEEE Int. Conf. Commun. Workshops (ICC Workshops)}, Jun. 2021, pp. 1-6.
\bibitem{IEEEhowto:kopka}
\label{ref:2023 Magazine_RSMA for 6G}
J. Park, B. Lee, J. Choi, H. Lee, N. Lee, S. H. Park, K. J. Lee, J. Choi,
S. H. Chae, S. W. Jeon, K. S. Kwak, B. Clerckx, and W. Shin, ``Rate-Splitting Multiple Access for 6G Networks: Ten Promising Scenarios and Applications,'' 2023, \emph{arXiv: 2306.12978}.
%\bibitem{IEEEhowto:kopka} \label{ref:Bruno_RSMA for Multi-Gateway}
%Z. W. Si, L. Yin, and B. Clerckx, "Rate-Splitting Multiple Access for Multigateway Multibeam Satellite Systems With Feeder Link Interference," \emph{IEEE Transactions on Communications}, vol. 70, no. 3, pp. 2147-2162, 2022.
%\bibitem{IEEEhowto:kopka} \label{ref:IoT_RSMA in SAIN}
%Z. Lin, M. Lin, T. de Cola, J.-B. Wang, W.-P. Zhu, and J. Cheng, "Supporting IoT With Rate-Splitting Multiple Access in Satellite and Aerial-Integrated Networks," \emph{IEEE Internet of Things Journal}, 2021.
%\bibitem{IEEEhowto:kopka} \label{ref:Arxiv_RSMAin GEO,LEO Co-Existing}
%W. U. Khan, Z. Ali, E. Langunas, S. Chatzinotas, and B. Ottersten, "Rate-Splitting Multiple Access for Cognitive Radio GEO-LEO Co-Existing Satellite Networks," \emph{arxiv:2208.02924v1}, 2022.
\bibitem{IEEEhowto:kopka} \label{ref:Arxiv_Bruno_RSMAinISTN}
L. Yin and B. Clerckx, ``Rate-Splitting Multiple Access for Satellite-Terrestrial Integrated Networks: Benefits of Coordination and Cooperation,'' \emph{IEEE Trans. Commun.}, vol. 22, no. 1, pp. 317-332, Jan. 2023.
%
\bibitem{IEEEhowto:kopka} \label{ref:Multicell-RSMA_1}
O. Tervo, L. N. Trant, S. Chatzinotas, B. Ottersten, and M. Juntti, ``Multigroup Multicast Beamforming and Antenna Selection with Rate-Splitting in Multicell Systems,'' in \emph{Proc. IEEE Int. Workshop Signal Process. Adv. Wireless Commun. (SPAWC)}, Jun. 2018, pp. 1-5.
\bibitem{IEEEhowto:kopka} \label{ref:Multicell-RSMA_2}
J. Zhang, J. Zhang, Y. Zhou, H. Ji, J. Sun, and N. Al-Dhahir, ``Energy and Spectral Efficiency Tradeoff via Rate Splitting and Common Beamforming Coordination in Multicell Networks,'' \emph{IEEE Trans. Commun.}, vol. 68, no. 12, pp. 7719-7731, Sep. 2020.

\bibitem{IEEEhowto:kopka} \label{ref:Multicell-RSMA_3}
X. Su, L. Li, and P. Zhang, ``Rate Splitting Based Robust Cooperative Transmission in Energy Cooperation Enabled Multi-Cell MISO System,'' in \emph{Proc. IEEE Wireless Commun, Netw. Conf. (WCNC)}, Apr. 2019, pp. 1-5.
\bibitem{IEEEhowto:kopka} \label{ref:Multicell-RSMA_4}
Y. Mao, B. Clerckx, and V. O. K. Li, ``Rate-Splitting Multiple Access for Coordinated Multi-Point Joint Transmission,'' in \emph{Proc. IEEE Int. Conf. Commun. Workshops (ICC Workshops)}, May 2019, pp. 1-6.
%\bibitem{IEEEhowto:kopka} \label{ref:Multicell-RSMA_5}
%\hl{N. Ha, W. Shin, M. Vaezi, and H. V. Poor, ``Coordinated Rate Splitting Multiple Access for Multi-Cell Downlink Networks,'' in \emph{Proc. Asilomar Conf. Signals, Syst., Comput.}, 2020, pp. 996-1001.}
%\bibitem{IEEEhowto:kopka} \label{ref:Multicell-RSMA_6}
%\hl{D. Yu, S. -H. Park, O. Simeone, and S. Shamai Shitz, ``Robust Design of Rate-Splitting Multiple Access With Imperfect CSI for Cell-Free MIMO Systems,'' in \emph{Proc. IEEE Int. Conf. Commun. Workshops (ICC Workshops)}, 2022, pp. 604-609.}
%\bibitem{IEEEhowto:kopka} \label{ref:Multicell-RSMA_4}
%\hl{Q. Zhu, Z. Qian, B. Clerckx, and X. Wang, ``Rate-Splitting Multiple Access in Multi-cell Dense Networks: A Stochastic Geometry Approach,'' \emph{arXiv:2207.11430v1}, 2022.
%}
\bibitem{IEEEhowto:kopka} \label{ref:Multicell-RSMA_5}
C. Gao, J. Zhang, L. Guo, K. Liu, L. Meng, and J. Sun, ``Coordinated rate splitting and power allocation in energy-spectral efficiency tradeoff-based multicell networks,'' \emph{Comput. Netw.}, vol. 218, Dec. 2022.

\bibitem{IEEEhowto:kopka} \label{ref:Han Kobayashi}
Te Han and K. Kobayashi, ``A new achievable rate region for the interference channel,'' \emph{IEEE Trans. Inf. Theory}, vol. 27, no. 1, pp. 49-60, Jan. 1981.
%\bibitem{IEEEhowto:kopka} \label{ref:Tse_Z channel}
%R. H. Etkin, D, N. C. Tse, and H. Wang, "Gaussian interference channel capacity to within one bit," \emph{IEEE Trans. Inf. Theory}, vol. 54, no. 12, pp. 5534-5562, 2008.
\bibitem{IEEEhowto:kopka} \label{ref:DVB-S2X}
``Second generation framing structure, channel coding and modulation systems for Broadcasting, Interactive Services, News Gathering and other broadband satellite applications; Part 2; DVB-S2 Extensions (DVB-S2X),'' Standard ETSI EN 302-307-2 V1.4.1, EBU, Feb. 2022.
%\emph{Second generation framing structure, channel coding and modulation systems for Broadcasting, Interactive Services, News Gathering and other broadband satellite applications; Part 2; DVB-S2 Extensions (DVB-S2X)}, Standard ETSI EN 302-307-2 V1.4.1, EBU, Feb. 2022.
%European Broadcasting Union (EBU)
\bibitem{IEEEhowto:kopka} \label{ref:HRS}
M. Dai, B. Clerckx, D. Gesbert, and G. Caire, ``A Rate Splitting Strategy for Massive MIMO With Imperfect CSIT,'' \emph{IEEE Trans. Wireless Commun.}, vol. 15, no. 7, pp. 4611-4624, Jul. 2016.
% \bibitem{IEEEhowto:kopka} \label{ref:BeyondDirtyCoding}
% Y. Mao and B. Clerckx, ``Beyond Dirty Paper Coding for Multi-Antenna Broadcast Channel With Partial CSIT: A Rate-Splitting Approach,'' \emph{IEEE Trans. on Commun.}, vol. 68, no. 11, pp. 6775-6791, Nov. 2020.
%\bibitem{IEEEhowto:kopka} \label{ref:RSMA_Survey}
%\hl{Y. Mao, O. Dizdar, B. Clerckx, R. Schober, P. Popovski and H. V. Poor, ``Rate-Splitting Multiple Access: Fundamentals, Survey, and Future Research Trends,'' \emph{IEEE Commun. Surv. Tutor.}, vol. 24, no. 4, pp. 2073-2126, 4th Quart., 2022.}
% Y. Mao, O. Dizdar, B. Clerckx, R. Schober, P. Popovski and H. V. Poor, "Rate-Splitting Multiple Access: Fundamentals, Survey, and Future Research Trends," in IEEE Communications Surveys & Tutorials, vol. 24, no. 4, pp. 2073-2126, Fourthquarter 2022
\bibitem{IEEEhowto:kopka} \label{ref:Math_Convex}
B. R. Marks and G. P. Wright, ``A General Inner Approximation Algorithm for Nonconvex Mathematical Programs,'' \emph{Open research}, vol. 26, no. 4, pp. 681-683, 1978.
\end{thebibliography}
\end{document}